\documentclass{article}

\usepackage{graphicx}
\usepackage{amsmath}
\usepackage[hidelinks]{hyperref}
\usepackage{float}
\usepackage{placeins}
\usepackage{comment}
\usepackage{color}

\begin{document}


\title{Coil-Integrated Alignment Sensor for Real-Time Feedback of Coil--Scalp Contact Point and Angle During Transcranial Magnetic Stimulation (TMS)}

\author{B. Seyed, M. Koehler, and S. M. Goetz}

\date{}

\maketitle

\begin{abstract}
Whereas coil positioning in transcranial magnetic stimulation (TMS) to reach a specific cortical target with modern focal stimulation coils has been intensively studied, the alignment and contact of a coil with the head is often ignored. Focal figure-of-eight coils have a point on the surface, where they generate the largest induced electric field. This point should touch the head first, and the coil should be approximately  tangential to the head in this point.
Previous research has demonstrated the large impact if the coil does not touch the head with the right point and that many operators struggle with establishing or maintaining the correct coil--scalp alignment.
This paper presents a technological support technology that can monitor the exact position of the contact point and also pressure to provide feedback to users. As the system uses exclusively components from consumer electronics, the sensor is low-cost and affordable. Through proper design, we achieved sufficient robustness so that the sensor does neither reset during TMS pulses and also not show any detectable degradation.

\end{abstract}

\section{Introduction}

Accurate coil placement remains one of the most operator-dependent steps in transcranial magnetic stimulation (TMS) \cite{Lin2022}. Small deviations in coil position and orientation can affect the induced electric field at the target and, consequently, stimulation accuracy and reproducibility \cite{Reijonen2020,Caulfield2022}. In routine practice, the challenge is not only to identify the intended target, but also to maintain a tangential alignment of the coil on the curved head and sufficient, but not excessive, pressure \cite{KoehlerAlignment}.

Current positioning solutions are expensive and still solve this problem only partially. Neuro\-navigation can improve  precision and reduce variability of coil positioning but does typically not solve the struggles of operators to align the coil tangentially \cite{Caulfield2022,Rodseth2017}.
When the coil is not perfectly tangential, a point other than the focus point with the strongest field touches the head. The focus point is lifted off though, which substantially reduces the effective stimulation in the target \cite{KoehlerAlignment, KoehlerAlignment2}.
This challenge appears to increase for curved and angled coils, although such coils have physical advantages and typically require less pulse energy for the same target field \cite{KoehlerCurvedCoils}.


As most targets outside the motor cortex do typically not provide acute quantifiable responses, there is no immediate feedback and alignment errors stay undetected.
Alignment errors often have more impact than misplacement due to the head curvature. A misalignment of a few degrees can already let a motor evoked potential vanish \cite{KoehlerAlignment}.

Poor control of the coil alignment may not just cause variability in outcomes. Thresholds found by inexperienced TMS users are often elevated due to alignment errors. As all subsequent procedures are relative to the threshold, the increased stimulation strength tends to increase pain for patients if rapid sequences such as theta burst and\,/\,or medial targets that co-activate the trigeminal nerve. Furthermore, the higher threshold readings may also affect safety of all repetitive TMS relative to motor threshold.

Without feedback and exclusive work on \textit{silent} targets outside the motor cortex, even experienced users can be unaware of the problem of coil alignment. In training clinical users for many years, the authors have observed that new users often experience difficulty particularly with the pitch of the coil and tend to touch the head with a point close to the coil handle first. For curved and angled coils, problems with the roll come on top.

Targets in neuronavigation are often already set with incorrect alignment so that the system's feedback is not helpful.
Many users are not aware or well acquainted with the four-dimensional cross-hair illustration of neuronavigation systems beyond the location. However, in any case, neuronavigation is often intrinsically not capable of managing the alignment accurately.
Without individual high-quality head scans and near-perfect scalp reconstruction, the estimates of the local surface normal and derived from it the alignment are anyway wrong \cite{zaman2023segmentation,KimReconstruction,Joshi1999}. However, the scalp reconstruction often demonstrates craters and artifacts. The alignment error estimation needs to calculate spatial derivatives based on the the noisy surface, which is numerically not well conditioned.

Robotic coil positioning may appear as the solution but faces the same alignment issues as the underlying neuronavigation system and human operators. \cite{Grosshagauer, RobotAccuracy}.


A solution should measure the alignment to guide operators through immediate online feedback and allow a combination with neuronavigation or also simpler 10--20 and cap-based positioning techniques that dominate clinical use \cite{Rajasekharan}. It could serve for training to raise awareness and intuition or, if practical enough, preferably for daily use. However, a solution would need to be robust, economical, and must not degrade the TMS performance. The latter implies that it should not add considerable weight to the coil and not excessively increase coil--cortex distance \cite{Stokes}.

We developed a coil-integrated sensing method based on a capacitive touch screen attached directly to the patient-facing side of the TMS coil. The sensor is interpreted by a computer with a custom graphical user interface (GUI). This system should not replace anatomical targeting methods but add immediate physical feedback about the coil--head interface and assist the operator in adjusting coil orientation during stimulation. 
The sensor can operate on its own to control coil--head alignment and contact. It can optionally be combined with conventional placement techniques, such as 10--20 grids on caps or endoscopic cameras in the focus point. Likewise, it can also complement conventional neuronavigation systems.

\section{Sensor Concept}
\begin{figure}[htbp]
\centering
\includegraphics[width=\textwidth,height=0.85\textheight,keepaspectratio]{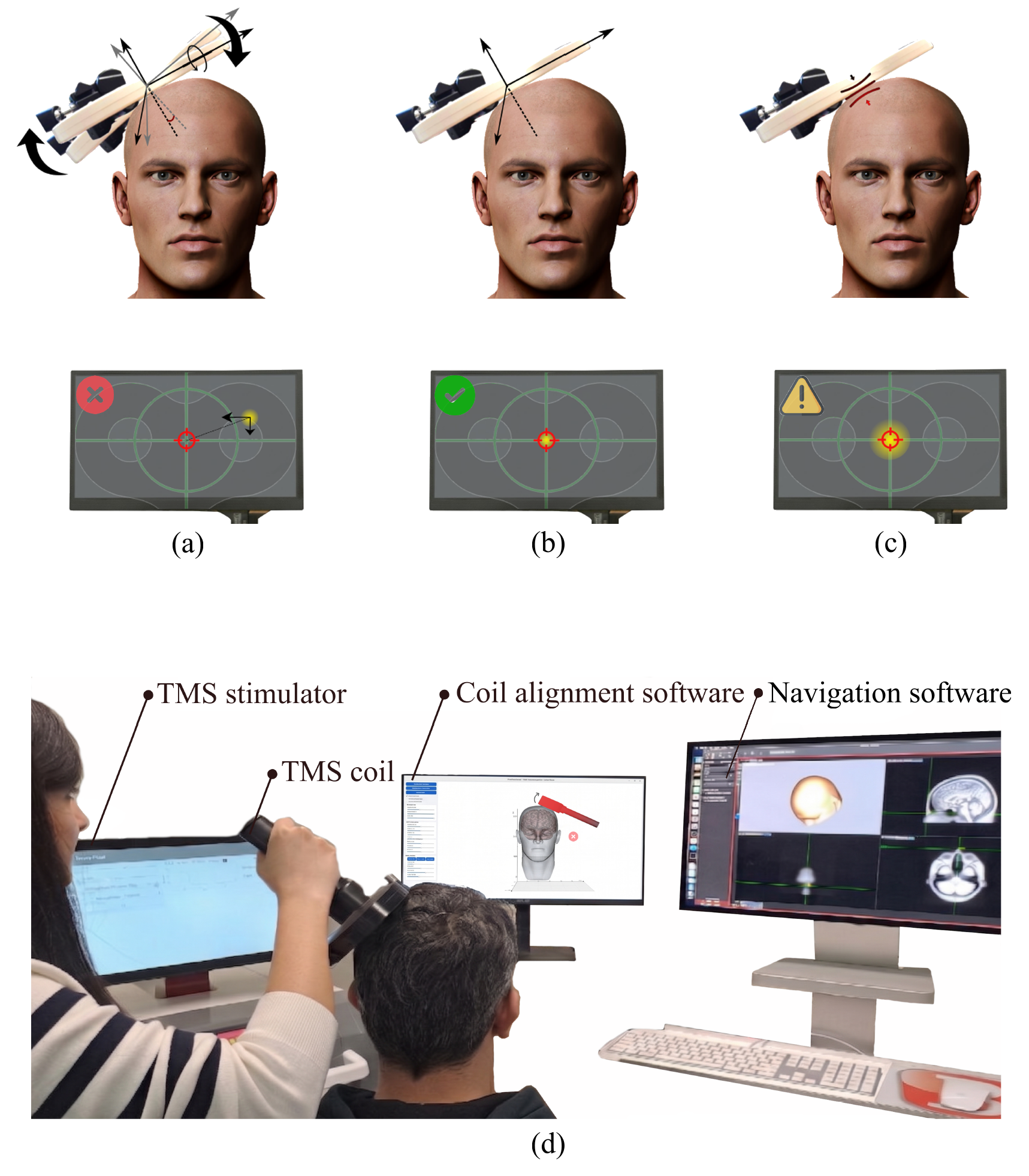}
\caption{Proposed contact-sensor feedback system for TMS coil positioning. (a)~Incorrect angle on the scalp, illustrated through a contact point on the touch sensor shifted away from the center. (b)~Coil positioned at the proper angle with a centered contact point on the sensor for good coil--scalp alignment. (c)~Additional illustration of the contact pressure and contact width through the size of the yellow cloud. (d)~Experimental setup in a TMS environment, here with an additional conventional neuronavigation system.}
\label{fig:tms_touch_sensor}
\end{figure}

Figure~\ref{fig:tms_touch_sensor} illustrates the operating concept. The coil is positioned over the patient's head above an intended cortical target. The capacitive contact sensor detects the two-dimensional position of the contact point through a grid of line and column traces that form mutual capacitances with the scalp. The sensor furthermore provides practical feedback on the contact pressure. The capacitive read-out furthermore isolates the skin contact and does not respond to the hair, e.g., in braids.

The contact sensor in its current form has a thickness of as low as 1.6~mm and is integrated in the bottom, patient-facing surface of the coil (Fig.~1(b)). 
This arrangement allows the sensing layer to be placed directly at the interface between the device and the scalp without substantially altering the operator's workflow. The sensor output is streamed to the GUI in real time and could detect the location of the touch point with a precision of { less than 1~mm in our implementation}.

\section{Electromagnetic Interference and Robustness}
A key engineering challenge was electromagnetic interference (EMI) from the powerful TMS pulses.
As the sensor is flat with a considerable area parallel to the winding, a large share of the magnetic flux permeates it and induces substantial electric fields across the surface, which the leads pick up and transport to the amplifier and read-out chip. Furthermore, the rapid  high-voltage swing of the TMS coil above (which is for most devices common mode, i.e., does not average out within the coil) further injects a strong capacitive artifact into the sensor electrodes relative to ground. The inductive and capacitive interference typically thwarts sensitive sensors at and near TMS coils.

We implemented several measures that finally enabled successful detection without failure or degradation of the sensor. We used wire routing of the flexible leads with minimum spanned area and guided the leads away from the interfering coil to interpret the line and column traces in detection electronics {(GT911, Goodix Technology, Shenzhen)} and transmit it serially to a computer (SOM10SG8B4CU, Singwon). 

We inserted a ferrite core to suppress high-frequency currents in the connection between the sensor and the detection electronics. Furthermore, we ground-lifted the entire signal processing and interpretation electronics and isolated its output towards the computer to suppress the capacitive interference of the TMS field pulse.

\section{Operation}
We tested the coil with more than 2,000 pulses and current rise times of $>150$~A/µs without any detectable degradation, reset, or change of behavior. The sensor furthermore continues to provide valid information in < 250~ms after a pulse so that users do not experience any interruption but assume continuous feedback.
The user interface (implemented in Python v3.14.3) illustrates the location of the contact point and the distance from the coil's focus point, where the coil should actually touch the scalp.

\section{Conclusion}

We presented a capacitive contact sensor mounted on the stimulation surface of a TMS coil that can provide real-time feedback on coil--head contact and rotational alignment that can manage the harsh electromagnetic interference of TMS with the described technical measures.

The sensor can work on its own but also offers a practical complement to existing TMS positioning workflows. Neuronavigation and robotic guidance are valuable for spatial targeting, whereas the present approach controls the proper alignment and contact as factors where previous approaches turned out insufficient or suboptimum. Because the hardware is compact and relatively simple to integrate, this approach may be useful in routine clinical sessions, operator training, and research in which consistency of coil--head interaction is important to achieve efficacy and reduce variability.

We are interested in improving the accuracy and reducing the variability of TMS as a concerted effort. In support of interested labs, we therefore would provide drawings and electronics upon request. In case of sufficient interest, we consider preparing a kit for retrofitting commercial coils.

%

\FloatBarrier

\pagebreak


\begin{thebibliography}{00}

\bibitem{Lin2022}
Y.-Y. Lin, R.-S. Chen, and Y.-Z. Huang (2022).
Impact of operator experience on transcranial magnetic stimulation.
Clinical Neurophysiology Practice, 7:42--48.

\bibitem{Reijonen2020}
J. Reijonen, L. S\"ais\"anen, M. K\"on\"onen, A. Mohammadi, and P. Julkunen (2020).
The effect of coil placement and orientation on the assessment of focal excitability in motor mapping with navigated transcranial magnetic stimulation.
Journal of Neuroscience Methods, 331:108521.

\bibitem{Caulfield2022}
K.A. Caulfield, H.H. Fleischmann, C.E. Cox, J.P. Wolf, and M.S. George (2022).
Neuronavigation maximizes accuracy and precision in TMS positioning: Evidence from 11,230 distance, angle, and electric field modeling measurements.
Brain Stimulation, 15:1192--1205.

\bibitem{KoehlerAlignment}
M. Koehler, T. Kammer, and S.M. Goetz (2022). How coil misalignment and mispositioning in transcranial magnetic stimulation affect the stimulation strength at the target. Clinical Neurophysiology, 162:159--161.

\bibitem{Rodseth2017}
J. Rodseth, E.P. Washabaugh, and C. Krishnan (2017).
A novel low-cost approach for navigated transcranial magnetic stimulation.
Restorative Neurology and Neuroscience, 35:601--609.

\bibitem{KoehlerAlignment2}
M. Koehler, T. Kammer, and S.M. Goetz (2023). Quantitative impact of coil misalignment and misplacement in transcranial magnetic stimulation. bioRxiv, 2023.11.18.567677. doi:10.1101/2023.11.18.567677 

\bibitem{KoehlerCurvedCoils}
M. Koehler and S.M. Goetz (2024). A closed formalism for anatomy-independent projection and optimization of magnetic stimulation coils on arbitrarily shaped surfaces. IEEE Transactions on Biomedical Engineering, 71(6):1745--1755.

\bibitem{Neuronavigation}
S.M. Goetz and T. Kammer (2021). Neuronavigation. Oxford Handbook of Transcranial Stimulation. 2nd edition, 183--226.



\bibitem{zaman2023segmentation}
F.A. Zaman, L. Zhang, H. Zhang, M. Sonka, and W. Wu (2023). Segmentation quality assessment by automated detection of erroneous surface regions in medical images, Computers in Biology and Medicine, 164:107324.

\bibitem{KimReconstruction}
S.H. Kim, Y.H. Choi, J.S. Lee, S.B. Lee, Y.J. Cho, S.H. Lee, S.-M. Shin, and J.-E. Cheon (2023). Deep learning reconstruction in pediatric brain MRI: comparison of image quality with conventional T2-weighted MRI. Neuroradiology, 65(1):207-214.

\bibitem{Joshi1999}
Mukta Joshi a, Jing Cui b, Keith Doolittle a, Sarang Joshi a, David Van Essen c, Lei Wang a, Michael I. Miller (1999). Brain segmentation and the generation of cortical surfaces. Neuroimage, 9(5):461-476.

\bibitem{Grosshagauer}
S. Grosshagauer, M. Woletz, M. Becher, J. Björklund, F. Padberg, D. Keeser, L. Bulubas, and C. Windischberger, C. (2025). Reducing target E-field variability in repetitive TMS through online motion compensation. Brain Stimulation, 19(1):102990.

\bibitem{RobotAccuracy}
S.M. Goetz, I.C. Kozyrkov, B. Luber, S.H. Lisanby, D.L.K. Murphy, W.M. Grill, and A.V. Peterchev (2019). Accuracy of robotic coil positioning during transcranial magnetic stimulation. Journal of Neural Engineering, 16(5):054003.

\bibitem{Stokes}
M.G. Stokes, C.D. Chambers, I.C. Gould, T.R. Henderson, N.E. Janko, N.B. Allen, and J.B. Mattingley (2005). Simple metric for scaling motor threshold based on scalp-cortex distance: application to studies using transcranial magnetic stimulation. Journal of Neurophysiology, 94(6):4520--4527.

\bibitem{Rajasekharan}
D. Rajasekharan, M.R. Madore, P. Holtzheimer, K.O. Lim, L.M. Williams, and N.S. Philip (2025). Personalized models of Beam\,/\,F3 targeting in transcranial magnetic stimulation for depression: Implications for precision clinical translation. Brain Stimulation, 18(3):829-837.

\end{thebibliography}
\end{document}